\documentclass[12pt]{article}
\usepackage{epsfig,cite}
\usepackage{amsmath}

\newcommand{\lsim}{\buildrel<\over{_\sim}}
\newcommand{\gsim}{\buildrel>\over{_\sim}}

\begin{document}
\thispagestyle{empty}

\def\thefootnote{\fnsymbol{footnote}}

\begin{flushright}
CERN--PH--TH/2004-101\hfill
DCPT/04/54\\
IPPP/04/27 \hfill
hep-ph/0406322
\end{flushright}

\vspace{1cm}

\begin{center}

{\large\sc {\bf LHC/LC Interplay in the MSSM Higgs Sector}}


 
\vspace{1cm}

{\sc 
K.~Desch$^{1}$%
\footnote{email: Klaus.Desch@desy.de}%
, E.~Gross$^{2}$%
\footnote{email: eilam.gross@weizmann.ac.il}%
, S.~Heinemeyer$^{3}$%
\footnote{email: Sven.Heinemeyer@cern.ch}%
, G.~Weiglein$^{4}$%
\footnote{email: Georg.Weiglein@durham.ac.uk}%
~and L.~\v{Z}ivkovi\'{c}$^{2}$%
\footnote{email: Lidiaz@wisemail.weizmann.ac.il}
}

\vspace*{1cm}

{\sl
$^1$ Institut f\"ur Experimentalphysik, Universit\"at Hamburg, 
Notkestr. 85,\\ 22607 Hamburg, Germany

\vspace*{0.4cm}

$^2$ Dept.\ of Particle Physics, Weizmann Institute of Science, 
Rehovot 76100, Israel

\vspace*{0.4cm}

$^3$ CERN, TH Division, Dept.\ of Physics, 1211 Geneva 23, Switzerland

\vspace*{0.4cm}

$^4$Institute for Particle Physics Phenomenology, University of Durham,\\
Durham DH1~3LE, UK
}

\end{center}

\vspace*{0.2cm}

\begin{abstract}
The interplay of prospective experimental information from both the 
Large Hadron Collider (LHC) and the Linear Collider (LC) 
in the investigation of the MSSM Higgs sector is analyzed in
the SPS~1a and SPS~1b benchmark scenarios. Combining LHC 
information on the heavy Higgs states of the MSSM with precise
measurements of the mass and branching ratios of the lightest CP-even
Higgs boson at the LC provides a sensitive consistency
test of the MSSM. This allows to set bounds on the trilinear
coupling $A_t$. In a scenario where LHC and LC only detect one light Higgs
boson, the prospects for an indirect determination of $M_A$ are investigated. 
In particular,
the impact of the experimental errors of the other SUSY parameters is
analyzed in detail. We find that a precision of about 20\% (30\%) can be
achieved for $M_A = $ 600 (800) GeV.
\end{abstract}

\def\thefootnote{\arabic{footnote}}
\setcounter{page}{0}
\setcounter{footnote}{0}

\newpage


\section{Introduction}

The prediction of a firm upper bound on the mass of the lightest Higgs
boson is one of the most striking predictions of Supersymmetric (SUSY)
theories
whose couplings stay in the perturbative regime up to a high energy
scale. Disentangling the structure of the Higgs sector and establishing
possible deviations from the Standard Model (SM) will be one of the
main goals at the next generation of colliders.

In order to implement electroweak symmetry breaking consistently into
the Minimal Supersymmetric Standard Model (MSSM), two Higgs doublets
are needed. This results in eight degrees 
of freedom, three of which are absorbed via the Higgs mechanism to give
masses to the $W^{\pm}$ and $Z$ bosons. The remaining five physical
states are the neutral CP-even Higgs bosons $h$ and $H$, the neutral
CP-odd state $A$, and the two charged Higgs bosons $H^{\pm}$. At the lowest
order, the Higgs sector of the MSSM is described by only two parameters
in addition to the gauge couplings, conventionally chosen as $M_A$ and
$\tan\beta$, where the latter is the ratio of the vacuum expectation
values of the two Higgs doublets.

The Higgs-boson sector of the MSSM is affected, however, by large
radiative corrections which arise in particular from the top and scalar
top sector and for large values of
$\tan\beta$ also from the bottom and scalar bottom sector.
Thus, the tree-level upper bound on the mass of the lightest CP-even Higgs
boson, $m_h < M_Z$ in the MSSM, arising from the gauge structure of the 
theory, receives large radiative corrections from the Yukawa sector of the 
theory~\cite{ERZ}. 
Taking corrections up to two-loop order into account, the mass is
shifted by about 50\%, establishing an upper bound of 
$m_h \lsim 136$~GeV~\cite{mhiggslong,Degrassi:2002fi}. 

An $e^+e^-$ Linear Collider (LC) 
will provide precision measurements of the properties of all
Higgs bosons that are within its kinematic reach~\cite{tesla,nlc,jlc}. 
Provided that a
Higgs boson couples to the $Z$~boson, the LC will observe it independently of
its decay characteristics. At the Large Hadron Collider (LHC), Higgs boson
detection can occur in various channels, see e.g.\ Ref.~\cite{HiggsLHC}. 
In many cases complementary 
information from more than one channel will be accessible at the LHC. In 
particular, the LHC has a high potential for detecting heavy Higgs
states which might be beyond the kinematic reach of the LC. 
Furthermore, experimental information on the parameters entering via
large radiative corrections will be crucial for SUSY Higgs
phenomenology. This refers in particular to a precise
knowledge of the top-quark mass, $m_t$, from the
LC~\cite{tesla,nlc,jlc,mtthreshold,deltamt} and information about the
SUSY spectrum  
from both LHC and LC~\cite{LHCLC}.

In the following, two examples of a possible interplay between LHC and
LC results in SUSY Higgs physics~\cite{LHCLC} are investigated. They
are based on the benchmark scenarios SPS~1a and SPS~1b~\cite{Allanach:2002nj}. 
In Section~\ref{sec:sps1b} a scenario is analyzed where the LHC can detect
the heavy Higgs states of the MSSM (see e.g.\ Ref.~\cite{Cavalli:2002vs}). 
This provides experimental information on both tree-level parameters of
the MSSM Higgs sector, $M_A$ and $\tan\beta$.
Therefore, in principle 
the phenomenology of the light CP-even Higgs boson can be predicted on
the basis of the experimental information on $M_A$ and $\tan\beta$.
The LC, on the other hand, provides precise 
information on the branching ratios of the light Higgs boson, which can
be compared with the theory prediction. 
A realistic analysis, however, requires to take into
account radiative corrections. In this way additional parameters become
relevant for predicting the properties of the light CP-even Higgs boson.
Comparing
these predictions with experimental results on the light CP-even Higgs
boson provides a sensitive consistency test of the MSSM at the quantum
level. This
allows in particular to obtain
indirect information on the mixing in the scalar top sector, which is
very important for fits of the SUSY Lagrangian to (prospective)
experimental data~\cite{sfittino}.
Deviations between the indirect predictions and the experimental results
may reveal physics beyond the MSSM. 

In Section~\ref{sec:sps1a} another scenario is analyzed where no
heavy Higgs bosons can be detected at LHC and LC.
The combined information about the SUSY spectrum from the LHC and LC and
of Higgs-boson branching ratio measurements at the LC is used to obtain
bounds on the mass of the CP-odd Higgs boson, $M_A$, in the
unconstrained MSSM (for such analyses within mSUGRA-like scenarios, see
Refs.~\cite{Dedes:2003cg,Ellis:2002gp}).
Since a realistic analysis requires the inclusion of radiative
corrections, the achievable sensitivity to $M_A$ depends on the
experimental precision of the additional input parameters and the
theoretical uncertainties from unknown higher-order corrections. This
means in particular that observed deviations in the properties of the
light CP-even Higgs boson compared to the SM case cannot be attributed
to the single parameter $M_A$. We analyze in detail the impact of 
the experimental and theory errors
on the precision of the $M_A$ determination. Our analysis
considerably differs from existing studies of Higgs boson
branching ratios in the literature~\cite{MAdet}.
In these previous analyses, all parameters except for the
one under investigation (i.e.\ $M_A$) have been kept fixed and the effect
of an assumed deviation between the MSSM and the SM has solely been attributed
to this single free parameter. This would correspond to a situation with
a complete knowledge of all SUSY parameters without any experimental or
theoretical uncertainty, which obviously leads to an unrealistic
enhancement of the sensitivity to the investigated parameter. 

Section~4 contains our conclusions.


\section{Scenario where LHC information on heavy Higgs states is
available}
\label{sec:sps1b}

In this section we analyze a scenario where experimental results 
at the LHC are used as input for confronting the predictions for the branching
ratios of the light CP-even Higgs boson with precision measurements at
the LC. We consider the SPS~1b benchmark
scenario~\cite{Allanach:2002nj}, which is a 
`typical' mSUGRA scenario with a relatively large value of $\tan\beta$. 
In particular, this scenario yields an $M_A$ value of about 550~GeV,
$\tan\beta = 30$, and stop and sbottom masses in the range of
600--800~GeV. 
More details about the mass spectrum
can be found in the Appendix and in Ref.~\cite{Allanach:2002nj}.

We assume the following experimental information from the LHC and the LC:
\begin{itemize}
\item
$\Delta M_A = 10\%$\\
This prospective accuracy on $M_A$ is rather conservative. The
assumption about the experimental accuracy on $M_A$ is not crucial in the
context of our analysis, however, since for $M_A \gg M_Z$ the
phenomenology of the light CP-even Higgs boson depends only weakly
on $M_A$.

\smallskip
\item
$\tan\beta > 15$\\
The observation of heavy Higgs states at the LHC in channels like 
$b \bar b H/A, H/A \to \tau^+\tau^-, \mu^+\mu^-$ will be possible in
the MSSM if  
$\tan\beta$ is relatively large~\cite{Cavalli:2002vs}. An attempted
determination of $\tan\beta$ 
from the comparison of the measured cross section with the theoretical
prediction will suffer from sizable QCD uncertainties, from the 
experimental errors of the SUSY parameters entering the theoretical
prediction, and from the experimental error of the measured cross
section.
Nevertheless, the detection of
heavy Higgs states at the LHC will at least allow to establish a lower
bound on $\tan\beta$. On the other hand, if $\tan\beta \lsim 10$ the LC
will provide a precise determination from measurements in the chargino
and neutralino sector. Thus, assuming a lower bound of $\tan\beta > 15$
seems to be reasonable in the scenario we are analyzing.

\item
$\Delta m_{\tilde t}, \Delta m_{\tilde b} = 5\%$\\
We assume that the LHC will measure the masses of the scalar top and
bottom quarks with $5\%$ accuracy. This could be possible if precise
measurements of parameters in the neutralino and chargino sector are
available from the LC, see Ref.~\cite{LHCLC}. On the 
other hand, the measurements
at the LHC (combined with LC input) will only loosely constrain the
mixing angles in the scalar top and bottom sectors.
Therefore we have not made any assumption about their values, but have
scanned over the whole possible parameter space (taking into account the
$SU(2)$~relation that connects the scalar top and bottom sector).
It should be noted that for the prospective accuracy on the scalar top
and bottom reconstruction at the LHC we have taken the results of
studies at lower values of $\tan\beta$ ($\tan\beta = 10$).
While the stop reconstruction should not suffer 
from the higher $\tan\beta$ value assumed in the present
study, sbottom reconstruction is more involved, see
Ref.~\cite{LHCLC}. We assume that the reconstruction of hadronic $\tau$'s 
from the decay $\chi^0_2 \to \tau^+ \tau^- \chi^0_1$ will be possible and the 
di-tau mass spectrum can be used for a mass measurement of the scalar 
bottom quarks at the 5\% level. This still has to be verified by experimental 
simulation.

In the scenario we are studying here the
scalar top and bottom quarks are outside the kinematic limit of the LC.

\item
$\Delta m_t = 0.1$~GeV\\
The top-quark mass (according to an appropriate short-distance mass
definition) can be determined from LC measurements at the $t \bar t$
threshold with an accuracy of 
$\Delta m_t \lsim 0.1$~GeV~\cite{tesla,nlc,jlc,mtthreshold}.
It should be kept in mind that the experimental error at the LHC, 
$\Delta m_t^{\rm LHC} \lsim $1--2~GeV, would induce a parametric
uncertainty in the $m_h$ prediction of $\Delta m_h^{m_t} \approx $1--2~GeV.

\item
$\Delta m_h = 0.5$~GeV\\
At the LC the mass of the light Higgs boson can be measured with an
accuracy of 50~MeV (whereas at the LHC a precision of 200--300~MeV
seems to be feasible). In order to account for theoretical uncertainties
from unknown higher-order corrections
we assume an accuracy of $\Delta m_h = 0.5$~GeV in this
study (which dominates over the experimental resolution at the LHC
and/or LC). This assumes that a considerable reduction of the present
uncertainty of about 3~GeV~\cite{Degrassi:2002fi} will be achieved until
the LC goes into operation.

\end{itemize}

The experimental information from the heavy Higgs and scalar quark
sectors that we have assumed above can be used to predict the branching
ratios of the light Higgs boson. Within the MSSM, the knowledge of these
experimental input quantities will significantly narrow down the
possible values of the light Higgs branching ratios. Comparing this
prediction with the precise measurements of the branching ratios carried
out at the LC will provide a very sensitive consistency test of the
MSSM.

\begin{figure}[htb!]
\vspace{6em}
\begin{center}
\epsfig{figure=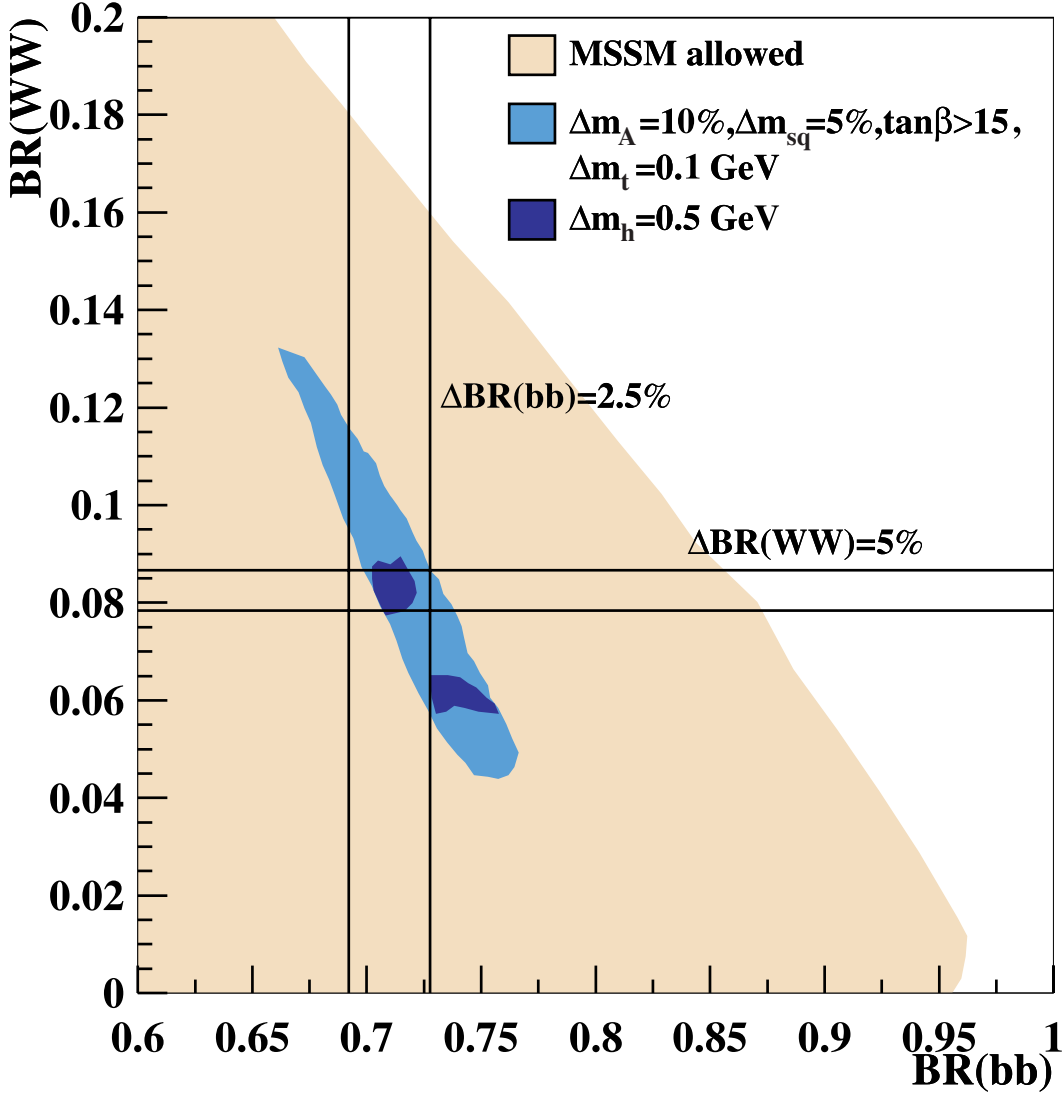, width=14cm}
\caption{
The experimental accuracies for the branching ratios BR($h \to b \bar b$) 
and BR($h \to WW^*$) at the LC of about 2.5\% and 5\%, indicated by a 
vertical and horizontal band, respectively, are
compared with the theoretical prediction in the MSSM. The light shaded
(yellow) 
region indicates the full allowed parameter space. The medium shaded
(light blue) region indicates the range of predictions in the MSSM
being compatible with the assumed experimental information from LHC
and LC,
$\Delta M_A = 10\%$, $\tan\beta > 15$,
$\Delta m_{\tilde t}, \Delta m_{\tilde b} = 5\%$,
$\Delta m_t = 0.1$~GeV.
The dark shaded (dark
blue) region arises if furthermore a measurement of the light CP-even
Higgs mass, including a theory uncertainty of $\Delta m_h = 0.5$~GeV,
is assumed. 
}
\label{fig:sec221plot}
\vspace{1em}
\end{center}
\end{figure}

This is shown in Fig.~\ref{fig:sec221plot} for the branching ratios
BR($h \to b \bar b$) and BR($h \to WW^*$). The light shaded (yellow) region
indicates the full parameter space allowed for the two branching ratios
within the MSSM. The medium shaded (light blue) region corresponds to
the range of predictions in the MSSM being compatible with the assumed 
experimental information from the LHC as discussed above, i.e.\
$\Delta M_A = 10\%$, $\tan\beta > 15$, 
$\Delta m_{\tilde t}, \Delta m_{\tilde b} = 5\%$. 
The shape and size of this region is mainly determined by the
experimental uncertainties on the top-quark mass and the scalar quark
masses; with the LC accuracy on $m_t$, the dominant uncertainty arises
from the scalar quark sector. The dark shaded (dark
blue) region arises if furthermore a measurement of the light CP-even
Higgs mass of $m_h = 116$~GeV, including a theory uncertainty of
$\Delta m_h = 0.5$~GeV, is assumed. 
The predictions are
compared with the prospective experimental accuracies for BR($h \to b
\bar b$) and BR($h \to WW^*$) at the LC of about 2.5\% and 5\%,
respectively~\cite{tesla,nlc,jlc,talkbrient}.

Agreement between the branching ratios measured at the LC and the
theoretical prediction
would constitute a highly non-trivial confirmation of the MSSM at the
quantum level. In
order to understand the physical significance of the two dark-shaded
regions in Fig.~\ref{fig:sec221plot} it is useful to investigate the
prediction for $m_h$ as a function of the trilinear coupling $A_t$ (see
also Ref.~\cite{deltamt}). If the masses of the scalar top and bottom quarks
have been measured at the LHC (using LC input), a precise measurement 
of $m_h$ will allow
an indirect determination of $A_t$ up to a sign ambiguity. It should be
noted that for this determination of $A_t$ the precise measurement of
$m_t$ at the LC is essential, see Fig.~3 in Ref.~\cite{deltamt}.%
\footnote{In Ref.~\cite{deltamt} it
was shown that going from $\Delta m_t = 2 (1)$~GeV to 
$\Delta m_t = 0.1$~GeV results in an improvement in the $A_t$
determination by a factor of 3 (2).
}%
~It also relies on a precise theoretical prediction for $m_h$, which
requires a considerable reduction of the theoretical uncertainties
from unknown higher-order corrections as compared to the present
situation~\cite{Degrassi:2002fi}, as discussed above. Making use of a 
prospective measurement of $m_h$ for predicting BR($h \to b \bar b$) and
BR($h \to WW^*$), on the other hand, is less critical in this respect,
since the kinematic effect of the Higgs mass in the prediction for the
branching ratios is not affected by the theoretical uncertainties.

\begin{figure}[htb!]
\begin{center}
\epsfig{figure=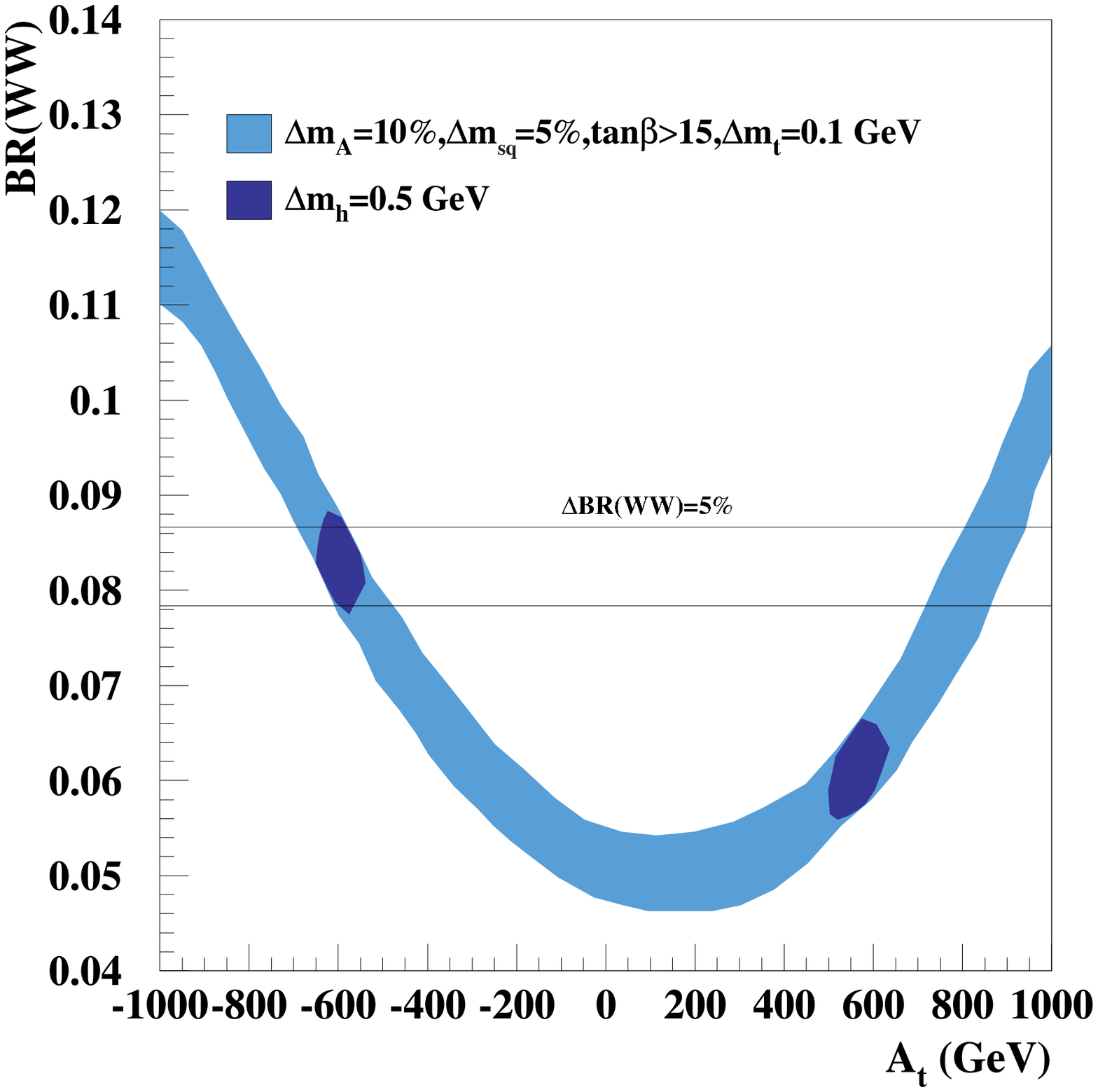, width=14cm}
\caption{
The branching ratio for $h \to WW^*$ is shown as a 
function of the trilinear coupling $A_t$. The light shaded (light blue)
region indicates the range of predictions in the MSSM
being compatible with the assumed experimental information,
$\Delta M_A = 10\%$, $\tan\beta > 15$,
$\Delta m_{\tilde t}, \Delta m_{\tilde b} = 5\%$,
$\Delta m_t = 0.1$~GeV. The dark shaded (dark
blue) region arises if furthermore a measurement of the light CP-even
Higgs mass, including a theory uncertainty of $\Delta m_h = 0.5$~GeV,
is assumed. The experimental
accuracy for BR($h \to WW^*$) at the LC of about 5\% is indicated by an
horizontal band.
}
\label{fig:sec221plotat}
\end{center}
\end{figure}

Fig.~\ref{fig:sec221plot} shows that the LC measurements of the
branching ratios of the light CP-even Higgs boson allow to discriminate
between the two dark-shaded regions. From the discussion above, these
two regions can be identified as corresponding to the two possible signs
of the parameter $A_t$. 
This is illustrated in Fig.~\ref{fig:sec221plotat},
where BR($h \to WW^*$) is shown as a function of $A_t$. It is demonstrated that
the sign ambiguity of $A_t$ can be resolved with the branching ratio
measurement. 
The determination of $A_t$ in this way will be crucial in 
global fits of the SUSY parameters to all available data~\cite{sfittino}.
As it is clear from Fig.~\ref{fig:sec221plotat}, this method of determining
the value and the sign of $A_t$ gets worse for small values of $|A_t|$.


\section{Indirect constraints on $M_A$ from LHC and LC\\
measurements}
\label{sec:sps1a}

In the following, we analyze an SPS~1a based scenario~\cite{Allanach:2002nj},
where we keep $M_A$ as a free parameter. We study in particular the
situation where the LHC only detects one light Higgs boson. For the
parameters of the SPS~1a scenario this corresponds to the region
$M_A \gsim 400$~GeV (see also the Appendix).

The precise measurements of Higgs branching ratios at the LC together
with accurate determinations of (parts of) the SUSY spectrum at the LHC
and the LC (see Ref.~\cite{LHCLC}) will allow in this
case to obtain indirect information on $M_A$ (for a discussion of
indirect constraints on $M_A$ from electroweak precision observables,
see Ref.~\cite{gigaz}). 
When investigating the sensitivity to $M_A$ it is crucial to take into
account realistic experimental errors of the other SUSY parameters that
enter the prediction of the Higgs branching ratios. 
Therefore we have varied all the SUSY
parameters according to error estimates for the measurements at
LHC and LC in this scenario. The sbottom masses and the gluino mass
can be obtained from mass reconstructions at the LHC with LC input,
see Ref.~\cite{LHCLC}. We have assumed a
precision of $\Delta m_{\tilde g} = \pm 8$~GeV and 
$\Delta m_{\tilde b_{1,2}} \approx \pm 7.5$~GeV.
We furthermore assume that the lighter stop (which in the SPS~1a
scenario has a mass 
of about 400~GeV, see Ref.~\cite{Allanach:2002nj}) will be accessible 
at the LC, leading to an accuracy
of about $\Delta m_{\tilde t_1} = \pm 2$~GeV. The impact of the LC information
on the stop mixing angle, $\theta_{\tilde t}$,
will be discussed below. For $\tan\beta$ we have 
used an uncertainty of $\Delta \tan\beta = 10\%$ (this accuracy can be 
expected from measurements at the LC in the gaugino sector for the SPS~1a 
value of $\tan\beta = 10$ \cite{Desch:2003vw}). As before, we have
assumed an error of $\Delta m_t = \pm 0.1$~GeV from the LC, so that
the parametric uncertainties on the $m_h$ predictions become negligible.
As above, we have
assumed a LC measurement of $m_h$, but included a theory error from
unknown higher-order corrections of $\pm
0.5$~GeV~\cite{Degrassi:2002fi}.

\begin{figure}[htb!]
\begin{center}
\epsfig{figure=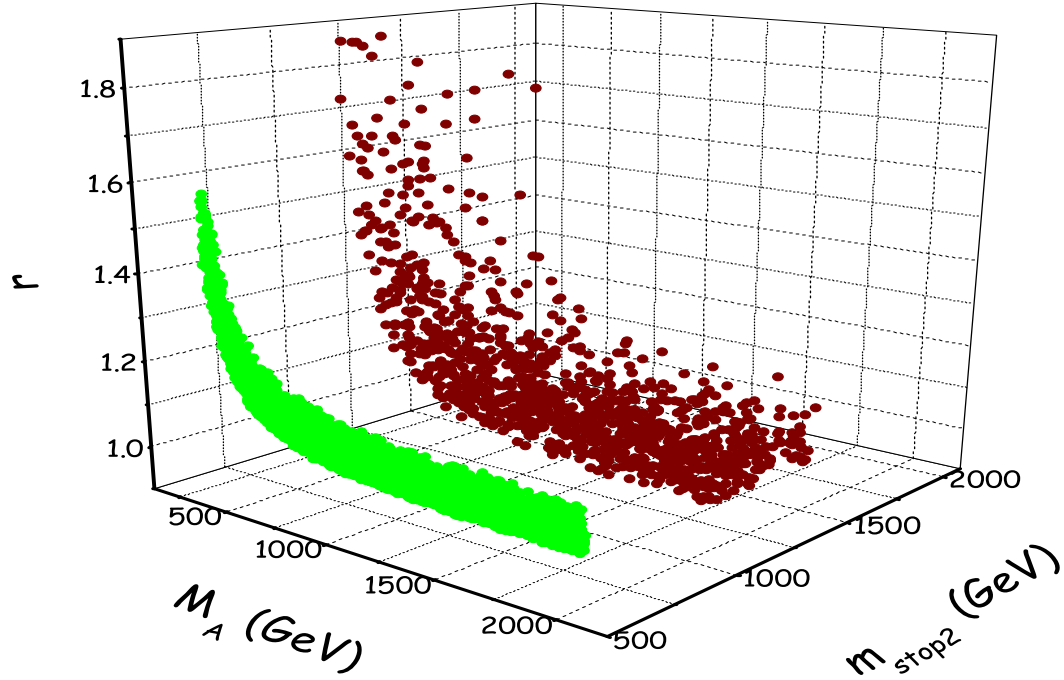, width=14cm,height=9.0cm}\\[3em]
\epsfig{figure=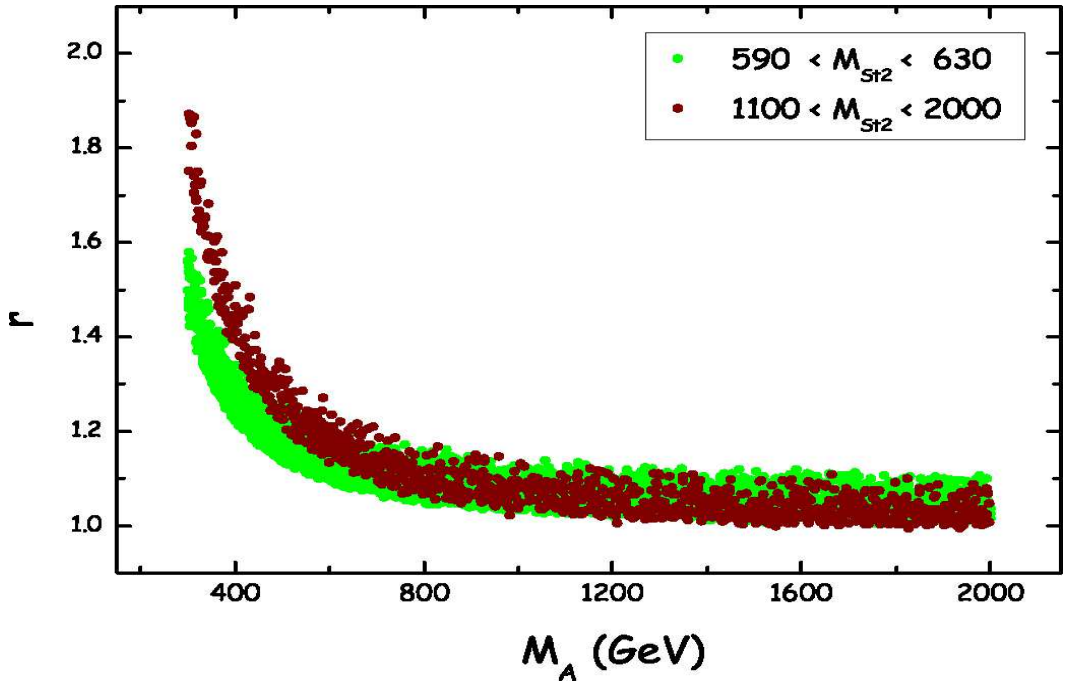, width=14cm,height=9.0cm}
\caption{
The ratio of branching ratios $r$, see eq.~(\ref{eq:sec221_r}),
is shown as a function of $M_A$ in the SPS~1a scenario. The
other SUSY parameters have been varied within the 3~$\sigma$ intervals
of their experimental errors (see text). The upper plot shows the 
three-dimensional $M_A$--$m_{\tilde t_2}$--$r$ parameter space, while
the lower plot shows the projection onto the $M_A$--$r$ plane.
}
\label{fig:sec221plotBR}
\end{center}
\end{figure}

\begin{figure}[htb!]
\begin{center}
\epsfig{figure=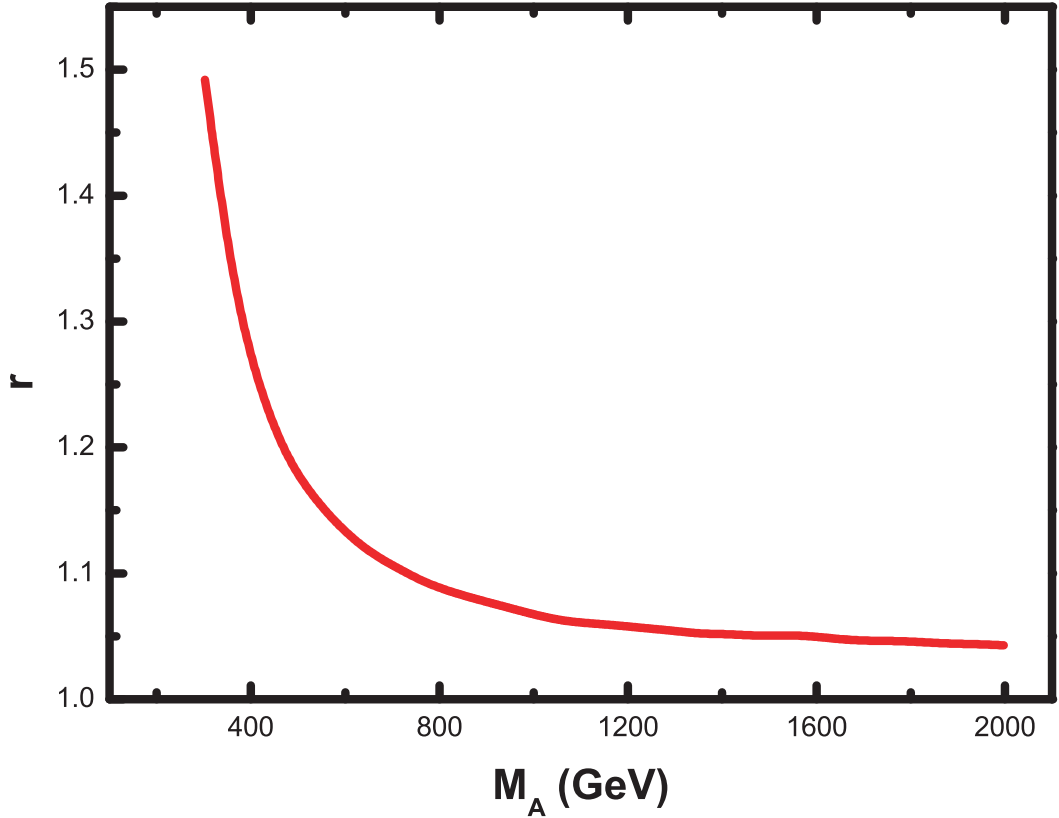, width=14cm}
\caption{The central value of $M_A$ corresponding to the central
  value of a prospective $r$~measurement is shown for the SPS~1a
  scenario. This relation between $r$ and $M_A$ would be obtained if all 
  experimental and theoretical uncertainties were negligible (see text).
}
\label{fig:rMA}
\end{center}
\end{figure}

In our analysis we compare the theoretical prediction~\cite{hff} for
the ratio of branching ratios 
\begin{equation}
r \equiv \frac{\left[{\rm BR}(h \to b \bar b)/
                     {\rm BR}(h \to WW^*)\right]_{\rm MSSM}}
              {\left[{\rm BR}(h \to b \bar b)/
                     {\rm BR}(h \to WW^*)\right]_{\rm SM~~~}\,}
\label{eq:sec221_r}
\end{equation}
with its prospective experimental measurement. 
Even though the experimental error on the ratio of the two BR's is
larger than that of the individual ones, the quantity $r$
has a stronger sensitivity
to $M_A$ than any single branching ratio.

In Fig.~\ref{fig:sec221plotBR} the theoretical prediction for~$r$
is shown as a function of $M_A$, where the scatter points result from
the variation of all relevant SUSY parameters within the 3~$\sigma$
ranges of their experimental errors. 
The constraint on the SUSY parameter 
space from the knowledge of $m_h$ is taken into account, where the
precision is limited by the theory uncertainty from unknown
higher-order corrections. The experimental information on $m_h$
gives rise in particular to indirect
constraints on the heavier stop mass and the stop mixing angle.%
\footnote{
Without the reduction of the intrinsic $m_h$ uncertainty and without a
precise determination of $m_t$ the constraint on the SUSY parameter
space would be much weaker, which would drastically decrease the
sensitivity to $M_A$.
}%
~Without
assuming any further experimental information, two distinct intervals
for the heavier stop mass (corresponding also to different intervals for
$\theta_{\tilde t}$) are allowed. This can be seen from the upper plot
of Fig.~\ref{fig:sec221plotBR}. The interval with lower values 
of $m_{\tilde t_2}$ corresponds to the SPS~1a scenario, while the
interval with higher $m_{\tilde t_2}$ values can only be realized in the
unconstrained MSSM. In the lower plot the projection onto
the $M_A$--$r$ plane is shown, giving rise to two bands with different
slopes. Since the lighter stop mass is accessible at the LC in this
scenario, it can be expected that the stop mixing angle will be
determined with sufficient accuracy to distinguish between the two
bands. This has an important impact on the indirect determination of
$M_A$. 

The central value of $r$ obtained from the band which is realized 
in the SPS~1a scenario is shown as a function of $M_A$ in
Fig.~\ref{fig:rMA}. The plot shows a non-decoupling behavior of $r$,
i.e.\ $r$ does not go to~$1$ for $M_A \to \infty$. This is due to the
fact that the SUSY masses are kept fixed in the SPS~1a scenario. 
In order to find complete decoupling, however, both $M_A$ and the mass
scale of the SUSY particles have to become large, see e.g.\
Ref.~\cite{decoupling}. It should be noted that the sensitivity of $r$
to $M_A$ is not driven by this non-decoupling effect. In fact, for
larger values of the SUSY masses the slope of $r(M_A)$ even increases 
(one example being the second band depicted in
Fig.~\ref{fig:sec221plotBR}). Thus, even stronger indirect bounds on
$M_A$ could be obtained in this case. 

\begin{figure}[htb!]
\begin{center}
\epsfig{figure=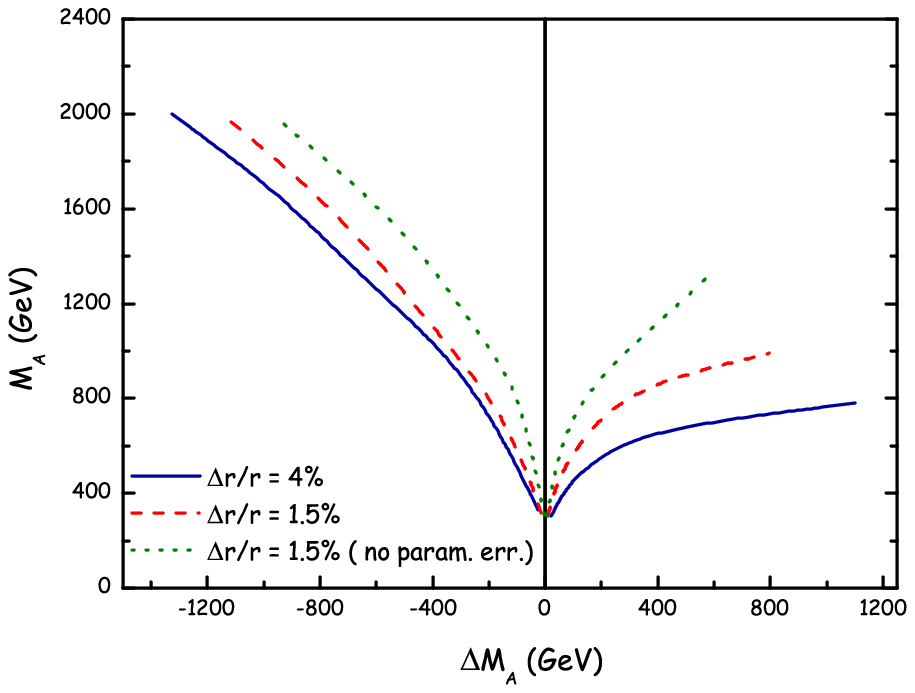, width=14cm}
\vspace{-0.5cm}
\caption{The 1~$\sigma$ bound on $M_A$, $\Delta M_A$,
versus $M_A$ obtained from a
comparison of the precision measurement of $r$ (see text) at the LC 
with the MSSM prediction. The results for $\Delta M_A$ are shown
for a 4\% accuracy of $r$ (full line) and a 1.5\% accuracy of $r$
(dashed line). The parametric uncertainties in the
prediction of $r$ resulting from LHC/LC measurement errors on
$\tan\beta, m_{\tilde{b}_{1,2}}, m_{\tilde{t}_1}, m_{\tilde{g}},
m_h $, and $m_t$ are taken into account. Also shown is the accuracy on
$M_A$ which would be obtained if these uncertainties were neglected 
(dotted line).
}
\label{fig:sec221plotBlackHole}
\vspace{-0.5cm}
\end{center}
\end{figure}

The relation between $r$ and $M_A$
shown in Fig.~\ref{fig:rMA} corresponds to an idealized situation where
the experimental errors of all input parameters in the prediction for
$r$ (besides $M_A$) and the uncertainties from unknown higher-order
corrections were negligibly small. 
The comparison of the theoretical prediction for $r$ (including
realistic uncertainties) with the experimental
result at the LC allows to set indirect bounds on the heavy Higgs-boson
mass $M_A$. 
Assuming a certain precision of $r$,
Fig.~\ref{fig:rMA} therefore allows to read off the best possible
indirect bounds on $M_A$ as a function of $M_A$, resulting from
neglecting all other sources of uncertainties. This idealized case is
compared with a more realistic situation based on the SPS~1a scenario in 
Fig.~\ref{fig:sec221plotBlackHole}.

For the experimental accuracy of $r$ we consider two
different values: a 4\% accuracy resulting from a first phase of LC
running with $\sqrt{s} \lsim 500$~GeV~\cite{tesla,nlc,jlc,talkbrient},
and a 1.5\% accuracy which can be achieved from LC running at 
$\sqrt{s} \approx 1$~TeV~\cite{barklow}.
In Fig.~\ref{fig:sec221plotBlackHole} the resulting 1~$\sigma$ 
bounds on $M_A$ are shown (the corresponding value of $r$ can be read off
from Fig.~\ref{fig:rMA}) for the experimental precisions of $r$ of 4\% 
and 1.5\%, respectively, where the estimated experimental errors on the
parameters 
$\tan\beta, m_{\tilde{b}_{1,2}}, m_{\tilde{t}_1}, m_{\tilde{g}},
m_h $, and $m_t$ based on the SPS~1a scenario 
are taken into account. Also shown is the 1~$\sigma$ error for 
$\Delta r/r = 1.5\%$ which would be obtained if all SUSY
parameters (except $M_A$) were precisely known, corresponding to the
idealized situation of Fig.~\ref{fig:rMA}. 

Fig.~\ref{fig:sec221plotBlackHole} shows that a 4\% accuracy on $r$ allows 
to establish an indirect upper bound on $M_A$ 
for $M_A$ values up to $M_A \lsim 800$~GeV (corresponding
to an $r$~measurement of $r \gsim 1.1$).
With an accuracy
of 1.5\%, on the other hand, a precision on $\Delta M_A / M_A$ of 
approximately 20\% (30\%) can be achieved for $M_A = $ 600 (800) GeV.
The indirect sensitivity extends to even higher values of $M_A$.
The comparison with the idealized situation where all SUSY parameters
(except $M_A$) were precisely known (as assumed in Ref.~\cite{MAdet})
illustrates the importance of taking 
into account the parametric errors as well as the
theory errors from unknown higher-order corrections. Detailed
experimental information on the SUSY spectrum and a precision
measurement of $m_t$ are clearly indispensable for exploiting the
experimental precision on $r$.


\section{Conclusions}

We have investigated indirect constraints on the MSSM Higgs and scalar
top sectors from
measurements at LHC and LC in the SPS~1a and SPS~1b benchmark scenarios.
In a situation where the LHC detects heavy Higgs bosons (SPS~1b) the
combination of the LHC information on the heavy Higgs states with
precise measurements of the mass and branching ratios of the lightest
CP-even Higgs boson at the LC gives rise to a sensitive consistency test
of the MSSM. In this way an indirect determination of the trilinear
coupling $A_t$ becomes possible. The measurement of $m_h$ alone allows
to determine $A_t$ up to a sign ambiguity, provided that a precise
measurement of the top-quark mass from the LC is available.
With the measurements of the branching ratios ${\rm BR}(h \to b \bar b)$ and
${\rm BR}(h \to WW^*)$ at the LC the sign ambiguity can be resolved and
the accuracy on $A_t$ can be further enhanced.

In a scenario where LHC and LC only detect one light Higgs
boson (SPS~1a, where $M_A$ is taken as a free parameter), 
indirect constraints on $M_A$ can be established from
combined LHC and LC data. Taking all experimental and theoretical
uncertainties into account, an indirect determination of $M_A$ with an
accuracy 
of about 20\% (30\%) seems to be feasible for $M_A = $ 600 (800) GeV.
In order to achieve this, a precise measurement of the branching ratios 
${\rm BR}(h \to b \bar b)$ and ${\rm BR}(h \to WW^*)$ at the LC and
information on the parameters of the scalar top and bottom sector from
combined LHC / LC analyses will be crucial.



\section*{Appendix}

Our numerical evaluation is based on the SPS~1a and 1b benchmark 
scenarios that have been defined in Ref.~\cite{Allanach:2002nj}.
The relevant parameters of the two benchmark scenarios are
given below (more details can be found in
Ref.~\cite{Allanach:2002nj}). They are given in the $\overline{\rm DR}$
scheme at the top squark mass scale. 
$m_{\tilde t_{L,R}}$ and $m_{\tilde b_{L,R}}$ denote the
diagonal soft SUSY-breaking parameters in the $\tilde t$ and
$\tilde b$ mass matrices, respectively. 

\begin{itemize}

\item SPS~1a
\begin{equation}
\begin{aligned}
m_{\tilde t_L} &= 495.9 \;{\rm GeV}, \quad
&m_{\tilde t_R} &= 424.8 \;{\rm GeV}, \quad
&A_t &=  -510.0\;{\rm GeV}, \\ 
m_{\tilde b_L} &=  m_{\tilde t_L}, \quad
&m_{\tilde b_R} &=  516.9\;{\rm GeV}, \quad
&A_b &= -772.7 \;{\rm GeV}, \\
m_{\tilde g}   &=  595.2\;{\rm GeV}, \quad
&M_2 &= 192.7 \;{\rm GeV}, \quad
&M_1 &= 99.1 \;{\rm GeV}, \\
\mu &=  352.4\;{\rm GeV}, \quad
&M_A &=  393.6\;{\rm GeV}, \quad
&\tan\beta &= 10.
\end{aligned}
\end{equation}

\item SPS~1b
\begin{equation}
\begin{aligned}
m_{\tilde t_L} &= 762.5 \;{\rm GeV}, \quad
&m_{\tilde t_R} &= 670.7 \;{\rm GeV}, \quad
&A_t &= -729.3 \;{\rm GeV}, \\ 
m_{\tilde b_L} &=  m_{\tilde t_L}, \quad
&m_{\tilde b_R} &= 780.3 \;{\rm GeV}, \quad
&A_b &= -987.4 \;{\rm GeV}, \\
m_{\tilde g}   &= 916.1 \;{\rm GeV}, \quad
&M_2 &= 310.9 \;{\rm GeV}, \quad
&M_1 &= 162.8 \;{\rm GeV}, \\
\mu &= 495.6 \;{\rm GeV}, \quad
&M_A &= 525.5 \;{\rm GeV}, \quad
&\tan\beta &= 30.
\end{aligned}
\end{equation}

\end{itemize}



\subsection*{Acknowledgements}
This work has been supported by the European Community's Human
Potential Programme under contract HPRN-CT-2000-00149 Physics at
Colliders, 
by the Benoziyo center for high
energy physics, and by the German Federal Ministry of
Education and Research (BMBF) within the Framework of the
German-Israeli Project Cooperation (DIP).
S.H. thanks the DESY theory division for hospitality in the final
stages of this work.



\end{document}